\documentclass[preprint2]{emulateapj}

\usepackage{graphicx,times,bm}
\graphicspath{{./fig/}{./png/}}

\newcommand{\EQ}{\begin{equation}}
\newcommand{\EN}{\end{equation}}
\newcommand{\EQA}{\begin{eqnarray}}
\newcommand{\ENA}{\end{eqnarray}}

\newcommand{\Eq}[1]{Eq.~(\ref{#1})}

\newcommand{\App}[1]{Appendix~\ref{#1}}

\newcommand{\Fig}[1]{Figure~\ref{#1}}

\newcommand{\Tab}[1]{Table~\ref{#1}}

\newcommand{\bra}[1]{\langle #1\rangle}

{}
{}
{}

{}
{}
\newcommand{\meanEMF}{\overline{\mbox{\boldmath ${\cal E}$}}{}}{}
{}
{}
{}
{}
{}
\newcommand{\meanBB}{\overline{\mbox{\boldmath $B$}}{}}{}
{}
{}
{}
{}
{}
{}
{}
\newcommand{\meanJJ}{\overline{\mbox{\boldmath $J$}}{}}{}

{}
{}
{}

\newcommand{\hatk}{\hat{k}}
\newcommand{\hatkk}{\hat{\bm{k}}}

{}

{}
{}

\newcommand{\NNN}{\hat{\bm{N}}}
\newcommand{\RRR}{\hat{\bm{R}}}
\newcommand{\TTT}{\hat{\bm{T}}}

\newcommand{\kk}{\bm{k}}

\newcommand{\uu}{\mbox{\boldmath $u$} {}}

\newcommand{\BB}{\mbox{\boldmath $B$} {}}

\newcommand{\SSS}{\mbox{\boldmath $S$} {}}
\newcommand{\AAA}{\mbox{\boldmath $A$} {}}

\newcommand{\nab}{\mbox{\boldmath $\nabla$} {}}

\newcommand{\ii}{{\rm i}}

\newcommand{\dd}{{\rm d} {}}

\def\hf{h_{\rm f}}
\def\hm{h_{\rm m}}

\def\kf{k_{\rm f}}

\def\etat{\eta_{\rm t}}

\def\half{{\textstyle{1\over2}}}

\newcommand{\W}{\,{\rm W}}

\newcommand{\uHz}{\,\mu{\rm Hz}}

\newcommand{\s}{\,{\rm s}}

\newcommand{\kms}{\,{\rm km/s}}

\newcommand{\Mm}{\,{\rm Mm}}
\newcommand{\Mx}{\,{\rm Mx}}

\newcommand{\AU}{\,{\rm AU}}

\newcommand{\pT}{\,{\rm pT}}
\newcommand{\yjgr}[3]{ #1, {J.\ Geophys.\ Res.,} {#2}, #3}

\newcommand{\yapj}[3]{ #1, {ApJ,} {#2}, #3}

\newcommand{\yapjl}[3]{ #1, {ApJ,} {#2}, #3}

\newcommand{\yana}[3]{ #1, {A\&A,} {#2}, #3}
\newcommand{\yanas}[3]{ #1, {A\&AS,} {#2}, #3}

\newcommand{\yssr}[3]{ #1, {Spa. Sci. Rev.,} {#2}, #3}

\newcommand{\ygrl}[3]{ #1, {Geophys.\ Res.\ Lett.,} {#2}, #3}

\newcommand{\yjfm}[3]{ #1, {J.\ Fluid Mech.,} {#2}, #3}

\newcommand{\ypf}[3]{ #1, {Phys.\ Fluids,} {#2}, #3}

\newcommand{\ypp}[3]{ #1, {Phys.\ Plasmas,} {#2}, #3}

\newcommand{\yaraa}[3]{ #1, {ARA\&A,} {#2}, #3}

\newcommand{\yprl}[3]{ #1, {Phys.\ Rev.\ Lett.,} {#2}, #3}

\newcommand{\ymn}[3]{ #1, {MNRAS,} {#2}, #3}

\newcommand{\ysph}[3]{ #1, {Solar Phys.,} {#2}, #3}

\newcommand{\ypre}[3]{ #1, {Phys.\ Rev.\ E,} {#2}, #3}

\newcommand{\yjour}[4]{ #1, {#2}, {#3}, #4}

\newcommand{\ybook}[3]{ #1, {#2} (#3)}
\newcommand{\yproc}[5]{ #1, in {#3}, ed.\ #4 (#5), #2}

\newcommand{\sana}[1]{ #1, {A\&A}, submitted}

\begin{document}
\title{Scale-dependence of magnetic helicity in the solar wind}

\author{Axel Brandenburg\altaffilmark{1,2},
Kandaswamy Subramanian\altaffilmark{3},
Andr\'e Balogh\altaffilmark{4,5}, \&
Melvyn L.\ Goldstein\altaffilmark{6}
}

\altaffiltext{1}{
NORDITA, AlbaNova University Center, Roslagstullsbacken 23,
SE-10691 Stockholm, Sweden \email{brandenb@nordita.org}
}\altaffiltext{2}{
Department of Astronomy, Stockholm University, SE-10691 Stockholm, Sweden
}\altaffiltext{3}{
Inter-University Centre for Astronomy and Astrophysics,
Post bag 4, Ganeshkhind, Pune 411 007, India \email{kandu@iucaa.ernet.in}
}\altaffiltext{4}{
International Space Science Institute, Hallerstrasse 6,
Bern CH-3012, Switzerland
}\altaffiltext{5}{
Space and Atmospheric Group, Blackett laboratory, Imperial College,
Prince Consort Road, London, United Kingdom \email{a.balogh@imperial.ac.uk}
}\altaffiltext{6}{
Code 673, NASA-Goddard Space Flight Center, Greenbelt, Maryland 20771, USA \email{melvyn.l.goldstein@nasa.gov}
}

\date{~$ $Revision: 1.114 $ $}


\begin{abstract}
We determine the magnetic helicity, along with the magnetic energy, at high latitudes
using data from the  {\it Ulysses} mission. The data set spans the time period from 
1993 to 1996. The 
basic assumption of the analysis is that the solar wind is homogeneous.
Because the solar wind speed is high, we follow the 
approach first pioneered by Matthaeus et al.\
(1982, Phys.\ Rev.\ Lett.\ 48, 1256) by which, under the assumption
of spatial homogeneity, one can use Fourier transforms of the magnetic
field time series to construct one-dimensional spectra of the magnetic
energy and magnetic helicity under the assumption that the Taylor frozen-in-flow hypothesis is valid. That is a well-satisfied assumption for the data used in this study. The magnetic helicity derives from the skew-symmetric terms of the
three-dimensional magnetic correlation tensor, while the symmetric terms of the tensor are used to determine the magnetic energy spectrum.
Our results show a sign change of magnetic helicity at
wavenumber $k\approx2\AU^{-1}$ (or frequency $\nu\approx2\uHz$)
at distances below $2.8\AU$ and at
$k\approx30\AU^{-1}$ (or $\nu\approx25\uHz$) at larger distances.
At small scales the magnetic helicity is positive at northern heliographic
latitudes and negative at southern latitudes.
The positive magnetic helicity at small scales is argued to be the result
of turbulent diffusion reversing the sign relative to what is seen
at small scales at the solar surface.
Furthermore, the magnetic helicity declines toward solar minimum in 1996.
The magnetic helicity flux integrated separately over one hemisphere
amounts to about $10^{45}\Mx^2/{\rm cycle}$ at large scales and to a 3 times
lower value at smaller scales.
\keywords{MHD -- Turbulence}
\end{abstract}


\section{Introduction}

Over the past 30 years there has been considerable activity in estimating
magnetic and current helicities of the Sun's magnetic field both at the
surface \citep{See90,Pev95,Bao99,Pev00} as well as in the solar wind
\citep{MGS82,RK94,RK96} using a variety of spacecraft, in particular
{\it Voyager 2}.
The early motivation of \citet{See90} was the connection with the
$\alpha$ effect in mean-field dynamo theory.
Subsequent work confirmed his early findings that
the current helicity has a negative sign in the northern hemisphere
and a positive in the southern.
This also agreed with expectations according to which current helicity
is a proxy for kinetic helicity \citep{Kei83}, which is known to be
negative for cyclonic events in the northern hemisphere and positive in
the southern.

Later work in connection with dynamo theory in periodic domains clarified
that a correspondence between kinetic and current helicities can only
be expected to hold at and below the scale of the energy-carrying eddies of the
turbulence, because at larger scales the signs of current and magnetic
helicities should reverse \citep{B01}.
This is connected with magnetic helicity evolution and the fact that
in $\alpha$ effect dynamos
magnetic helicity at large and small scales tend to have opposite signs
\citep{See96,Ji99}.
If magnetic helicity fluxes and resistive effects are weak or unimportant,
e.g., in the kinematic regime of a growing dynamo, the total magnetic
helicity is constant or zero if it was zero initially.
The production of magnetic helicity at the scale of the energy-carrying
eddies is then accompanied by the production of magnetic and current
helicity of the opposite sign at scales larger than the scale of the
energy-carrying eddies.
There is a similar tendency also when total magnetic helicity is not
conserved, e.g., on long timescales when resistive effects become
important or if magnetic helicity fluxes are present.
Thus, there are good theoretical reasons to expect that the magnetic
field in the Sun is bi-helical, i.e., of opposite sign at small and
large length scales \citep{BB03}.

We emphasize that `small' refers here to the scale of the
energy-carrying eddies, which is also called the outer scale---in contrast
to the inner scale where kinetic and magnetic energies get dissipated.
In the Sun the outer scale can be as large as $50\Mm$, which corresponds
to the pressure scale height near the bottom of the solar convection zone.
On the other hand, `large' refers to the scale of the mean field
that characterizes the solar cycle.
The width of the toroidal flux belts, i.e., the width of the wings
of the butterflies in a solar butterfly diagram is about $20^\circ$,
corresponding to about $200\Mm$ in the Sun.
In any case, the scale of the large-scale field is finite, i.e., such
a field is still subject to decay and possible regeneration by dynamo
action, and should thus not be confused with an `imposed' magnetic field.
The latter case is sometimes considered in numerical simulations, where
the departure from an imposed field corresponds to the small-scale field
whose helicity is not conserved on its own \citep{SMO95,Ber97,BM04}.
Let us also mention at this point that the magnetic helicity is quadratic
in the magnetic field, so it is not expected to flip sign from one cycle
to the next, although it may of course vary in strength.

The bi-helical nature of the magnetic field has been the topic of
related work by \cite{YB03}, who investigated the relaxation of
an initially bi-helical field and the mutual annihilation of the two signs of
magnetic helicity.
It should be noted that a connection has also been discussed between
the current helicity observed in the Sun and that obtained from mean-field
dynamo models \citep{See90,DG01}.
Furthermore, \cite{CCN04} find mostly negative current helicity in the
north, except that during short intervals at the beginning of each cycle
the current helicity in the north can be positive, while
\cite{Zhang06} find a band of negative current helicity at mid-latitudes
and positive values at higher and lower latitudes.
However, these papers ignored the possibility that the magnetic field
in each hemisphere is expected to be bi-helical.
The first observational evidence for a reversed sign of magnetic helicity at
large scales came from an analysis of synoptic maps of the radial
magnetic field of the Sun \citep{BBS03}.
They found a sign reversal of magnetic helicity at the time of
solar maximum with positive values after that moment.

There is yet another reason for studying magnetic helicity in the Sun.
Magnetic helicity is a topological invariant that is equal to half the number
of constructive flux rope crossings times the square of the magnetic
flux in these ropes \citep{Mof69}.
Magnetic helicity is therefore a measure of the degree of tangledness.
It might then be possible to assess the degree of tangledness by counting
the net crossings of filaments in ${\rm H}_\alpha$ images of the Sun
\citep{Chae00}, although this technique leaves some ambiguity regarding
the sign of the magnetic helicity.
Other ways of measuring magnetic helicity and their fluxes is by tracking
the motions at the solar surface \citep{Kus02,DB03},
which led to the estimate that the
total flux of magnetic helicity in each hemisphere, integrated over a
full 11-year cycle, is of the order of $10^{46}\Mx^2/{\rm cycle}$ \citep{BR00}.
This number agrees also with theoretical expectations of an upper
limit of this value \citep{BS04,Bra09}.
However, there is no indication as to a possible scale dependence
of the magnetic helicity.
The only evidence for this is just the qualitative appearance of a
systematic tilt of bipolar regions.
This tilt corresponds to writhe helicity, which is a quantity that
depends only on the topology of the axis of a flux tube structure.
Independent of the time during the 11-year cycle, it should have
a positive sign in the northern hemisphere and a negative sign
in the southern hemisphere, i.e., just the opposite of what is observed
in the magnetic field line twist at smaller scales.

The hope is now that measurements of the solar wind might help teach
us something about the scale dependence of the contribution to the
magnetic field that is related to the $\alpha$ effect.
In order for the $\alpha$ effect to work efficiently and to escape what
is known as catastrophic $\alpha$ quenching, negative magnetic helicity
associated with small-scale magnetic fields must be shed \citep{BF00,KMRS00};
see \cite{BS05} for a review.
This might be accomplished by coronal mass ejections \citep{BB03}.
Coronal mass ejection events are manifold,
and they are almost all associated with magnetic helicity \citep{Dem02},
but concern only the corona and not the solar wind.
The large-scale magnetic field in the solar wind is characterized by the
Parker spiral \citep{Par58}.
The helicity associated with the Parker spiral is known to be negative
in the northern hemisphere and positive in the southern \citep{Bie87} --
independent of the time during the 11-year cycle.
Therefore, even though we would normally associate the Parker spiral with the
large-scale field, its helicity is of opposite sign to the helicity of the large-scale
field generated in the dynamo interior, or that expected from the tilt
of the flux tubes near the solar surface. On the other hand, the sign of the helicity
associated with the small-scale field that needs to be shed does agree with that of the 
Parker spiral, although it would seem to be of the wrong scale.
Nevertheless, not much is known about the relationship between
magnetic helicity fluxes and magnetic helicity itself.
For example, it is possible that turbulence in the region outside the
dynamo would continue to diffuse the magnetic field, although it would
no longer amplify it by an $\alpha$ effect.
This effect would tend to reverse the production of bi-helical magnetic
fields and would pump positive magnetic helicity into smaller scales,
leaving behind negative magnetic helicity at larger scales.
This could be interpreted as a forward turbulent cascade, but it is
probably only possible in an expanding flow, so as to not cause a conflict
with the realizability condition that enforces the inverse transfer
in a confined helical flow.
This phenomenon was seen in mean-field calculations with a turbulent exterior
\citep[see Fig.~7 of][]{BCC09} and also in direct numerical simulations
of dynamos in spherical geometry with a nearly force-free exterior
\citep[see Fig.~4 of][]{WBM10}.

To probe quantitatively the possible scale dependence
of the magnetic field, perhaps the best type of analysis is that used in the
early measurements on {\it Voyager 2}.
Making use of the Taylor hypothesis, \cite{MGS82} were able to
associate frequencies with wavevectors.
Making the further assumption of homogeneity, they were able to translate the
simultaneous measurement of the two field components perpendicular to the
direction of the wind to information not only about the magnetic energy
spectrum, but in particular, about the magnetic helicity spectrum.
The background for application of this technique to the computation of other
helicities was explored further in \cite{MGS86}.

A possible complication with many of the early results is that
the trajectories of {\it Voyager} and
other spacecraft were close to the ecliptic, across which the magnetic helicity
is expected to change sign. However, there are no published spectra that
appear to involve data sets that crossed the heliospheric current sheet.
Nevertheless, the magnetic helicity as found by \cite{MG82} 
randomly changed sign at all scales, although 
\cite{GRF91} did find short intervals during which the magnetic helicity
had a constant sign at scales close to that of the proton gyroradius.
Furthermore, \cite{SB93} found that at frequencies below $10\uHz$ the
magnetic helicity tends to have a predominant sign: negative in the north
and positive in the south, while at higher frequencies there are strong
fluctuations of the sign.
However, we shall argue below that, by averaging over broad wavenumber
bins, it is still possible to extract meaningful information from the
data even at high frequencies.
The technique applied by \cite{MGS82} appears very
suitable for the purpose of assessing scale dependence of magnetic helicity.
The purpose of the present work is therefore to apply this technique
to more recent measurements of {\it Ulysses}
that flew in a nearly polar orbit that covered both hemispheres.

\section{Data analysis}

We use $60\s$ time averages from the Vector Helium Magnetometer on
{\it Ulysses}.
The original time resolution is up to 2 vectors/second and the sensitivity
is about $10\pT$; see \cite{Bal92} for a detailed description.
The available data comprise measurements of all three components of the
magnetic field $\BB$ and velocity $\uu$ in the locally Cartesian heliospheric coordinate
system $(R,T,N)$, where $R$ is the distance from the Sun, $T$ points in the
transverse direction parallel to the solar equatorial plane and is
positive in the direction of solar rotation, and $\NNN=\RRR\times\TTT$
is the third direction pointing toward heliographic north.
This corresponds to a right-handed coordinate system.
Note that the $R$-$T$ plane is inclined to the heliographic equatorial
plane by an angle equal to the heliographic latitude $\lambda$ of the
spacecraft.

We have analyzed 27 data sets comprising a time span of about one month each
and covering different epochs between 1993 and 1996.
During 1993/94, {\it Ulysses} was at $39^\circ$ to $80^\circ$ southern
latitudes and distances between $4.3\AU$ to $1.7\AU$, while during 1995/96
it was at $43^\circ$ to $79^\circ$ northern latitudes and distances
between $1.5\AU$ to $3.5\AU$.
%
%
For each data set we determined the average radial wind speed $u_R$,
which ranges between $720$ and $790\kms$.
Note that the local wind speed points almost exactly in the direction
away from the Sun.
Using Taylor's hypothesis, we translate time $t$ into the negative
radial coordinate, $R=R_0-u_R t$, where $R_0$ is the slowly changing
distance of the spacecraft.
Next, we compute the Fourier transform of each of the field components,
\EQ
\tilde{B}_i(k_R)=\int e^{\ii k_R R} B_i(R)\,\dd R,
\quad i=R,T,N,
\EN
and are thus able to compute the spectral correlation matrix,
\EQ
M_{ij}^{\rm1D}(k_R)=\tilde{B}_i(k_R) \tilde{B}_j^*(k_R),
\label{M1D}
\EN
where the asterisk denotes complex conjugation.
The superscript 1D emphasizes an important difference
with the three-dimensional correlation tensor $M_{ij}^{\rm3D}(\kk)$.
Before discussing this in more detail we note that,
in practice, $M_{ij}^{\rm1D}(k_R)$ is obtained from measurements
along the $R$ direction.
In that case one computes the one-dimensional magnetic energy and
helicity spectra simply as $\mu_0 E_{\rm M}^{\rm1D}(k_R)=|\hat{\BB}|^2$
and $H_{\rm M}^{\rm1D}(k_R)=4\,{\rm Im}(\hat{B}_T\hat{B}_N^\star)/k_R$.
These are the equations used by \cite{MGS82}, and we
shall use them for the most part of this paper as well.

It is well known that no assumption about isotropy is made in obtaining
the one-dimensional magnetic energy and helicity spectra.
However, we should emphasize that, even if the turbulence were isotropic,
the one-dimensional spectra obtained through direct measurements are
{\it not} equivalent to the three-dimensional ones \citep{TL72}.
The differences can become important in regions where the spectra
deviate from pure power law scaling \citep{Dobler_etal03}.
The purpose of the rest of this section is to extend the
well-known formula for the conversion between
one- and three-dimensional energy spectra to the case of helicity spectra.
To highlight the analogy between the two, we ignore here
the consideration of longitudinal and transverse energy spectra and
make the assumption of isotropy.
This assumption does seem at odds with results obtained in the ecliptic
\citep[see, e.g.,][]{NSGG10,Sah10}, but is consistent with an analysis of
{\it Ulysses} data reported by \cite{Smi03}.
Even though there is near isotropy of the variances
(shown also below), there is no spectral
isotropy, so the correlation length perpendicular to $\BB$ is
shorter than along $\BB$.
However, by making the assumption of isotropy, we shall be able to assess
the differences between three- and one-dimensional spectra.
It will turn out that these differences are rather small.

In three-dimensional isotropic helical turbulence we have
\EQ
M_{ij}^{\rm3D}(\kk)=(\delta_{ij}-\hatk_i\hatk_j)\,{2\mu_0 E_{\rm M}^{\rm3D}(k)\over8\pi k^2}
-\epsilon_{ijk}{\ii k_kH_{\rm M}^{\rm3D}(k)\over8\pi k^2},
\label{IsotropicTensor}
\EN
where $\hatkk=\kk/k$ is the unit vector of $\kk$.
Note that these spectra obey the realizability condition,
\EQ
2\mu_0 E_{\rm M}^{\rm3D}(k)\ge k|H_{\rm M}^{\rm3D}(k)|,
\EN
where the factor $2$ in front of $E_{\rm M}(k)$ is just a consequence
of the factor 1/2 in the definition of energy.
The two spectra are normalized such that
\EQ
\int\delta_{ij}M_{ij}^{\rm3D}(\kk)\,\dd^3k=
\int_0^\infty \mu_0 E_{\rm M}^{\rm3D}(k)\,\dd k=\bra{\BB^2}/2,
\EN
\EQ
\int\epsilon_{ijl}\frac{\ii k_l}{k^2} M_{ij}^{\rm3D}(\kk)\,\dd^3k=
\int_0^\infty H_{\rm M}^{\rm3D}(k)\,\dd k=\bra{\AAA\cdot\BB},
\EN
where $\BB=\nab\times\AAA$ is the magnetic field expressed in terms
of the magnetic vector potential $\AAA$, which obeys the Coulomb gauge,
$\nab\cdot\AAA=0$, and angular brackets denote averaging over the
data spanned by each data set.

Let us now relate $M_{ij}^{\rm3D}(\kk)$ to $M_{ij}^{\rm1D}(k_R)$.
Suppose $M_{ij}^{\rm3D}(\kk)$ were known, then,
to improve the statistics, one can obtain $M_{ij}^{\rm1D}(k_R)$ from
$M_{ij}^{\rm3D}(\kk)$ by averaging over the other two wavevector
components, i.e.,
\EQ
M_{ij}^{\rm1D}(k_R)=\int M_{ij}^{\rm3D}(k_R,k_T,k_N)\,\dd k_T\,\dd k_N.
\label{Relation1D3D}
\EN
We define one-dimensional energy and helicity spectra via
\EQ
\delta_{ij}M_{ij}^{\rm1D}(k_R)=\mu_0 E_{\rm M}^{\rm1D}(k_R),
\label{EM1D}
\EN
\EQ
2\epsilon_{ijR}(\ii k_R)^{-1} M_{ij}^{\rm1D}(k_R)=H_{\rm M}^{\rm1D}(k_R),
\label{HM1D}
\EN
and note that they are related to the three-dimensional energy and
helicity spectra via an integral transformation
\EQ
E_{\rm M}^{\rm1D}(k_R)=\int_{k_R}^\infty E_{\rm M}^{\rm3D}(k)\,\dd\ln k,
\label{EMto1D}
\EN
\EQ
H_{\rm M}^{\rm1D}(k_R)=\int_{k_R}^\infty H_{\rm M}^{\rm3D}(k)\,\dd\ln k.
\label{HMto1D}
\EN
This transformation is well known for the energy spectrum
\citep[cf.,][]{TL72,Dobler_etal03}, but has been generalized here
to the case with helicity; see \App{Deriv} for details of the derivation.
In the following we present results first for 
$E_{\rm M}^{\rm1D}$ and $H_{\rm M}^{\rm1D}$ and
compute then the three-dimensional spectra via differentiation, i.e.,
\EQ
E_{\rm M}^{\rm3D}(k)=-\dd E_{\rm M}^{\rm1D}(k)/\dd\ln k,
\EN
\EQ
H_{\rm M}^{\rm3D}(k)=-\dd H_{\rm M}^{\rm1D}(k)/\dd\ln k.
\EN
Note, however, that differentiation amplifies the error
in the already rather noisy data.
Therefore we perform differentiation based on data that have
been averaged into rather broad wavenumber bins.
We use a second-order midpoint formula with respect to $\ln k$ bins.
This yields
data at $k$ values that lie between those for the one-dimensional spectra.
In the following, where we present mostly one-dimensional spectra, we omit
the superscript 1D, but retain superscript 3D for all three-dimensional
spectra.

\section{Results}

\begin{figure}[t!]\begin{center}
\includegraphics[width=\columnwidth]{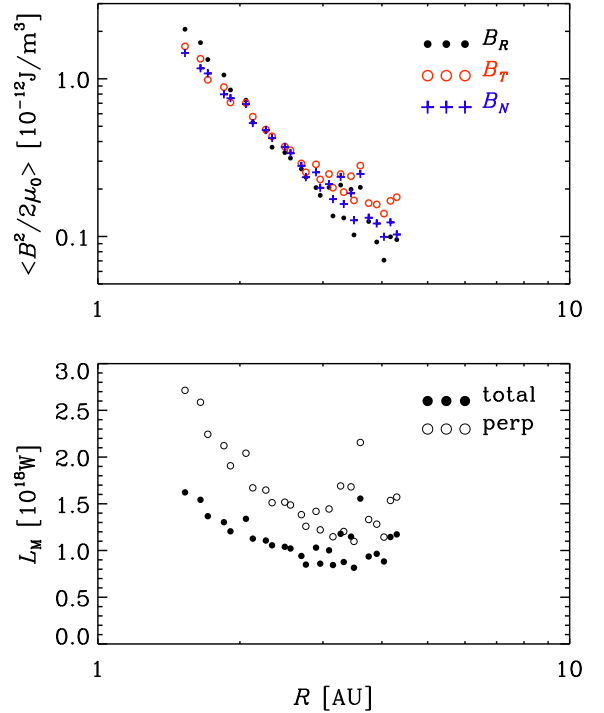}
\end{center}\caption[]{
Magnetic energy density and magnetic luminosity as a function of distance.
Note that the decay of magnetic energy is slightly faster than
$\propto R^{-2}$ and that the magnetic `luminosity' varies in the range
$4$--$8\times10^{17}\W$.
}\label{pdistance}\end{figure}

For the 27 data sets analyzed, the distance of the spacecraft to the
Sun varies from $\sim1.5\AU$ to $\sim4.5\AU$.
This needs to be taken into account when combining different data sets.
In \Fig{pdistance} we show that the integrated magnetic energy in the
three field components is approximately equal
in the parameter range covered by the 27 data sets.
This is compatible with earlier results of \cite{MGS86b} using data from
{\it Voyager 2}.
Furthermore, all three contributions fall off slightly
faster with distance than $\propto R^{-2}$.
One would expect a perfect $R^{-2}$ scaling if the Poynting flux stayed
constant, which one might expect for a magnetically dominated wind.
This suggests that magnetic energy is dissipated into heat, which has
been discussed in detail in recent years
\citep{GRM95,TM95,Fre98,Smith01,Sah09,Sah10}.
An estimate for the corresponding magnetic `luminosity', assuming approximate isotropy
over the solid angle, i.e.,
\EQ
L_{\rm M}=\oint(\BB^2/2\mu_0)\,\uu\cdot\dd\SSS
=4\pi R^2 \bra{\BB^2/2\mu_0}\,u_R,
\EN
also falls off from $2.8\times10^{18}\W$ at $1.5\AU$ to
$1.2\times10^{18}\W$ at $4.5\AU$.
The magnetic luminosity based on the perpendicular field components
shows a weaker decline from $1.6$ to $0.8\times10^{18}\W$.
This might be a consequence of the Parker spiral for which one expects
a $B_T/B_R\propto R$ scaling \citep{Bie87,Webb10}.
The present data suggest that this is however a weak effect.
Our proxy for the magnetic luminosity corresponds to (2--$4)\times10^{-9}L_\odot$,
where $L_\odot$ is the bolometric luminosity of the Sun.
Thus, because of the approximate $R^{-2}$ dependence we scale the
spectra $E_{\rm M}(k)$ and $H_{\rm M}(k)$ to a reference distance of $1\AU$
before averaging over different data sets.

\begin{figure}[t!]\begin{center}
\includegraphics[width=\columnwidth]{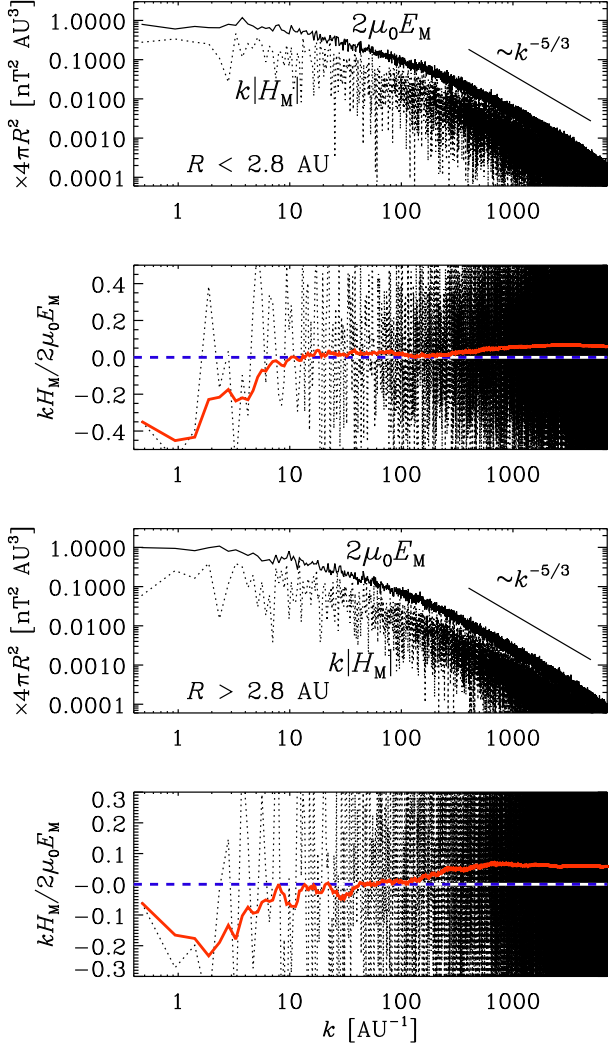}
\end{center}\caption[]{
Magnetic energy and helicity spectra, $2\mu_0 E_{\rm M}(k)$
and $k|H_{\rm M}(k)|$, respectively, for two separate distance intervals
(first and third panels).
Furthermore, both spectra are scaled by $4\pi R^2$ before averaging
within each distance interval above and below $2.8\AU$, respectively.
The relative magnetic helicity, $kH_{\rm M}(k)/2\mu_0 E_{\rm M}(k)$,
is plotted separately (second and fourth panels) together with its
cumulative average starting from the low wavenumber end.
The zero line is shown in dashed.
}\label{rmeanhel4b_EH}\end{figure}

Next, we focus on the magnetic energy and helicity spectra.
We begin by presenting results where we combine data from both hemispheres.
As we will verify later, the magnetic helicity has opposite signs
in the northern and southern hemispheres,
so we multiply the helicity measured in the south by $-1$.
Another possibility would be to divide by $\sin\lambda$, but at least for the
Parker spiral one does expect a much sharper sign change near the
equator than what is expected from a $\sin\lambda$ profile \citep{Bie87}.
Furthermore, we distinguish between data sets where the distance to
the Sun is either inside or outside $2.8\AU$.
In \Fig{rmeanhel4b_EH} we plot, separately for two separate distance
intervals, $2\mu_0 E_{\rm M}(k)$ and $k|H_{\rm M}(k)|$, rescaled by
$4\pi R^2$, as well as the relative magnetic helicity,
$kH_{\rm M}(k)/2\mu_0 E_{\rm M}(k)$.
We also show a cumulative average of this ratio,
starting from the low wavenumber end.
This shows quite clearly that the magnetic helicity in the north is negative
at small wavenumbers (large length scales) and that it becomes positive
at large wavenumbers (small length scales).
The cumulative nature of the average slightly overemphasizes the negative
contributions, delaying thus the position where the sign changes.
The negative magnetic helicity at large scales agrees with that belonging to
the Parker spiral, while the positive magnetic helicity at small scales
could be the result of turbulent diffusion leading to a reverse transfer
of magnetic helicity from large to small scales, as explained in the introduction.

\begin{figure}[t!]\begin{center}
\includegraphics[width=\columnwidth]{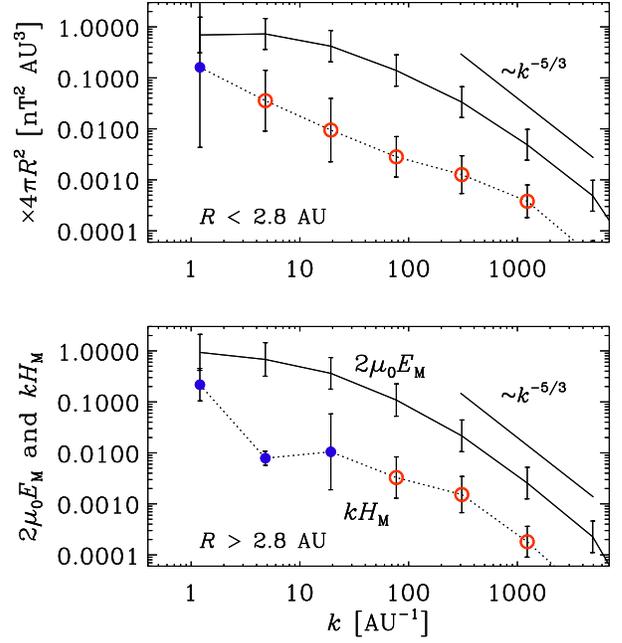}
\end{center}\caption[]{
Magnetic energy and helicity spectra, $2\mu_0 E_{\rm M}(k)$
and $kH_{\rm M}(k)$, respectively, for two separate distance intervals.
Furthermore, both spectra are scaled by $4\pi R^2$ before averaging
within each distance interval above and below $2.8\AU$, respectively.
Filled and open symbols denote negative and positive values of
$H_{\rm M}(k)$, respectively.
}\label{rmeanhel4b}\end{figure}

In the remainder of this paper we use logarithmically spaced wavenumber bins.
This way we can reduce the data to a small number of bins
and thereby minimize the statistical noise in a more meaningful way.
This is shown in \Fig{rmeanhel4b}.
In each bin we average the actual values (not the logarithms)
of spectral energy and magnetic helicity, both weighted with a $k$ factor.
We use rather broad bins, for example the first wavenumber bin at
$k=1.5\AU^{-1}$ has a width of $\Delta k=1.2\AU^{-1}$, and the second bin at
$k=5.8\AU^{-1}$ has $\Delta k=4.8\AU^{-1}$.
Given a maximum length of 1 month for each of the 27 data sets,
we have in principle a spectral resolution of $\Delta\Omega=0.2\dd^{-1}$,
corresponding to $\Delta k=0.5\AU^{-1}$.
We use data that were already averaged over $60\s$ time intervals.
This corresponds to a spatial resolution of $50\Mm$ and hence a Nyquist
wavenumber of $10^4\AU^{-1}$.
At distances below $2.8\AU$, the magnetic helicity is negative only in
the smallest wavenumber bin ($k\approx1\AU^{-1}$), while at distances
beyond $2.8\AU$ the first 3 wavenumber bins ($k<30\AU^{-1}$) show negative
helicity.

An estimate for the error has been obtained by comparing with the
averages that result by taking only data from the northern or the
southern hemisphere into account.
The larger one of the two departures is taken as an estimate of the error.
This results in a relative uncertainty of our average values by a
factor of 1--2.

\begin{figure}[t!]\begin{center}
\includegraphics[width=\columnwidth]{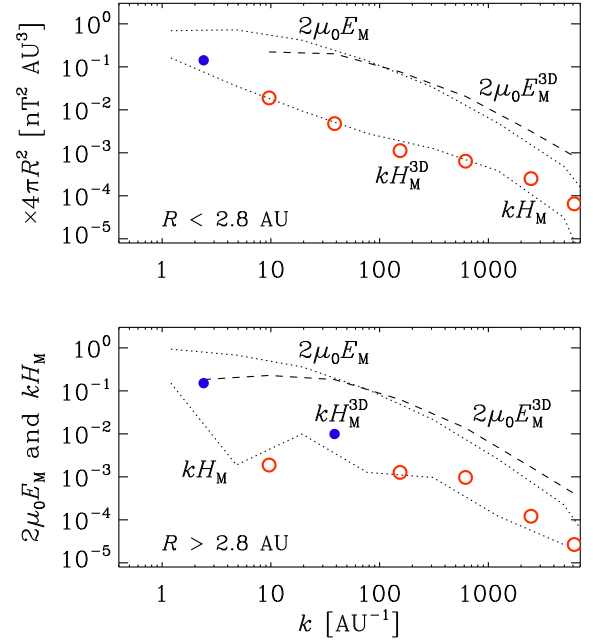}
\end{center}\caption[]{
Comparison between 3D and 1D spectra of magnetic energy and helicity.
In both panels the 1D spectra are denoted by dotted lines, while the 3D
spectra of magnetic energy by dashed and the 3D spectra of magnetic
helicity are indicated by filled and open symbols for negative and
positive contributions.
}\label{rmeanhel4b_3D}\end{figure}

As explained above, three-dimensional spectra of magnetic energy and
magnetic helicity can be obtained by differentiation with respect to
$\ln k$.
It turns out that, within expected error margins, the three-dimensional
spectra are surprisingly close to the one-dimensional spectra.
This is shown in \Fig{rmeanhel4b_3D} where we compare the types of spectra.
In view of the fact that the three-dimensional spectra do not seem to
alter our conclusions, and since differentiation
increases the noise in the data, we restrict ourselves in the following
to the discussion of one-dimensional spectra.

It is worthwhile noting the large separation between both graphs in the ordinate.
In other words, $k|H_{\rm M}(k)|\ll2\mu_0 E_{\rm M}(k)$.
This indicates that the relative magnetic helicity is rather small, which
may not be too surprising considering the fact that we have averaged
a noisy magnetic helicity result over rather broad wavenumber bins.
Also, of course, there is no reason to expect the relative magnetic helicity
at the solar surface to be particularly strong.
Next, from the open and filled symbols we can see the sign of $H_{\rm M}(k)$,
which turns out to be negative at the largest scales (filled symbols) and positive
at small scales (open symbols).
At a larger distance from the Sun, the break point where the sign of the
magnetic helicity changes grows to larger wavenumbers, corresponding to
smaller scales.
This is probably again related to the effect of turbulent diffusion
causing the reversed transfer of magnetic helicity from larger to
progressively smaller scales.
It could also be a consequence of the growing dominance of the large-scale
field (having negative helicity) with distance, so that it would appear as if the magnetic helicity
of the large-scale field imprints itself onto the smaller scales.
A similar effect has been seen in helical dynamo simulations with
an imposed magnetic field \citep[see Fig.~3 of][]{BM04}, where for
sufficiently strong fields the sign of the magnetic helicity is equal
to that of the kinetic helicity at small scales.
At the same time, the relative magnetic helicity diminishes, which is
indeed also seen in \Fig{rmeanhel4b}.

\begin{figure}[t]\begin{center}
\includegraphics[width=\columnwidth]{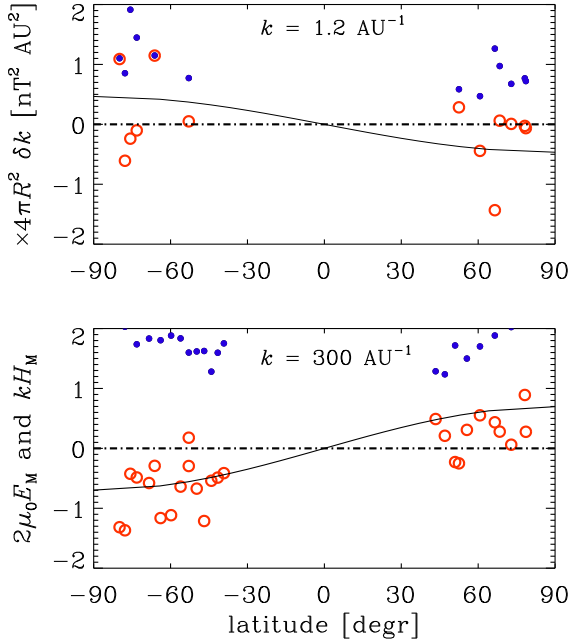}
\end{center}\caption[]{
Magnetic helicity (open symbols) and energy (filled symbols)
vs.\ latitude for data from wavenumber bin
$k=1.2\AU^{-1}$ and distances $R<2.8\AU$ (upper panel) and wavenumber
bin $k=300\AU^{-1}$ and all distances (lower panel).
In the last panel the energy is downscaled by factor 10, because
otherwise those data points would lie outside the plot range.
The solid lines represent fits proportional to $\sin\lambda$,
while dash-dotted lines represent the zero value.
}\label{plat4b}\end{figure}

In view of dynamo theory and for comparison with earlier work, it is
of interest to compute magnetic helicity fluxes separately for large
and small length scales, and integrate them over half the solid angle
and over the 11 year cycle, $T_{\rm cyc}$, i.e., we compute
\EQ
\half{\cal L}_{\rm H}^{\pm}T_{\rm cyc}
=2\pi R^2 u_R T_{\rm cyc} \int_{k\stackrel{<}{>} \kf}H_{\rm M}(k)\,\dd k,
\EN
where the 1/2 factor takes the fact into account that one normally
gives magnetic helicity fluxes integrated separately for each hemisphere
\citep{BR00}.
In \Tab{Tsummary} we give the results for
$\half{\cal L}_{\rm H}^{\pm}T_{\rm cyc}$ separately for
large ($-$) and small ($+$) scales and also for small and large distances.
It turns out that these values are typically around $10^{45}\Mx^2/{\rm cycle}$,
which is remarkably close to early estimates of \cite{BR95} of
$2\times10^{45}\Mx^2/{\rm cycle}$, and
about 10 times below the expected upper limit \citep{Bra09}.

\begin{table}[b]\caption{
Results for $\half{\cal L}_{\rm H}^{\pm}T_{\rm cyc}$ in units of
$\Mx^2/{\rm cycle}$.
}\vspace{12pt}\centerline{\begin{tabular}{ccc}
Distance & \quad Large Scales \quad & \quad Small Scales \quad \cr
\hline
$R<2.8\AU$ & $-0.9\times10^{45}$ & $+0.3\times10^{45}$  \\
$R>2.8\AU$ & $-1.3\times10^{45}$ & $+0.03\times10^{45}$  \\
\label{Tsummary}\end{tabular}}\end{table}

\begin{figure}[t]\begin{center}
\includegraphics[width=\columnwidth]{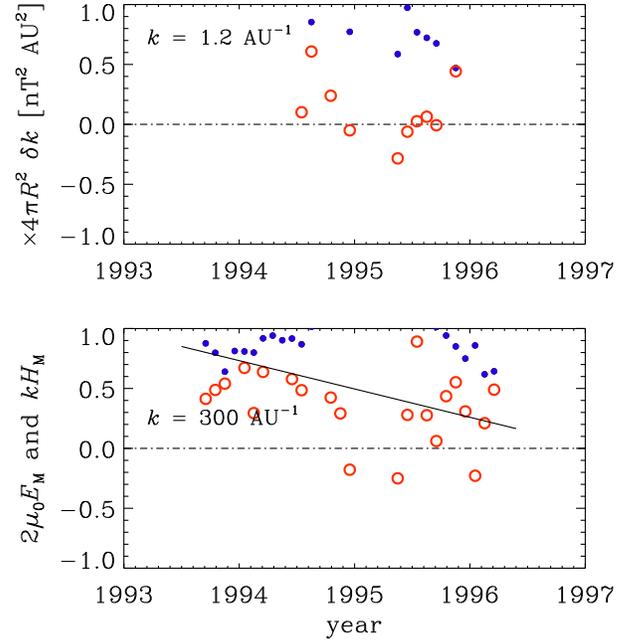}
\end{center}\caption[]{
Similar to \Fig{plat4b}, but vs.\ time.
Again, in the last panel the energy is downscaled by factor 10, because
otherwise those data points would lie outside the plot range.
The dash-dotted lines represent the zero value.
}\label{ptime4b}\end{figure}

Next, we consider the latitudinal dependence by abandoning the
averaging over heliographic latitude and consider data from two separate
wavenumber bands around 1.2 and $300\AU^{-1}$.
The data are obviously very noisy now, especially at low wavenumbers
where the wavenumber bins involve fewer data points; see \Fig{plat4b}.
Nevertheless, there is still some
evidence for the magnetic helicity having opposite signs in the two
hemispheres and, in addition, opposite sign at large ($5\AU$) and
small ($0.02\AU\approx3000\Mm$) length scales ($=2\pi/k$).

The present data have all been taken from the time just after solar
maximum and before the next solar minimum.
It would therefore be interesting to compare this with measurements
taken at other times.
There is in principle the possibility that the magnetic helicity
could change and even reverse sign for brief time intervals.
To see whether already the present data suggest a possible
systematic temporal trend, we plot the spectral magnetic helicity in
the same two wavenumber bins (around 1.2 and $300\AU^{-1}$)
vs.\ time; see \Fig{ptime4b}.
Given the high noise level, it is not possible to conclude anything
definite from the plot.
Only at large wavenumbers (shown here for $k=300\AU^{-1}$) there
might be a meaningful downward trend.
Of course, one can only be really sure about this when comparing data
over a much longer time span, in which case one might expect to see an
oscillatory variation as is also seen in Figures 3 and 4 of \cite{BR00}.

\section{Conclusions}

The present results have revealed for the first time evidence that the
magnetic helicity in the solar wind has opposite signs at large and
small length scales.
However, the signs are actually the other way around than what was
naively predicted based on the expected signs at the solar surface
\citep{BB03}, but they do agree with more recent simulations of
\cite{BCC09} and \cite{WBM10,WBM11} that show a reversed sign some distance away
from the dynamo
regime due to the effect of turbulent diffusion (without $\alpha$ effect)
that tends to forward-cascade magnetic helicity from large to small scales.

Although the solar wind
would carry some imprint of the fields generated within the Sun,
this field would be deformed due to the fact that the
material leaves the Sun with a net angular momentum, and
is perhaps turbulent. 
The magnetic helicity generation by turbulence 
may be understood using the magnetic
helicity conservation equation for the gauge invariant small-scale
helicity $\hf$ derived by \cite{SB06}.
We have already used this argument in this paper when we associated the
effect of turbulent magnetic diffusion with a forward turbulent cascade.
Suppose that the wind is turbulent on small scales,
then the helicity generation on those scales is governed by 
$\partial\hf/\partial t=-2\meanEMF\cdot\meanBB+...$,
where $\meanEMF$ is the mean turbulent electromotive force and the dots
refer to microscopic dissipation and possible flux terms.
As in mean-field dynamo theory \citep[see, e.g.,][]{BS05}, we approximate 
$\meanEMF\sim\alpha\meanBB-\etat\meanJJ$, where
$\meanJJ=\nab\times\meanBB$ is the mean current density,
and $\alpha$ and $\etat$ refer to a possible $\alpha$ effect
and turbulent diffusion associated with solar wind turbulence,
then $\partial\hf/\partial t=-2\alpha\meanBB^2+2\etat\meanBB\cdot\meanJJ$.
The dynamo generated large-scale field $\meanBB$, 
which has positive helicity, will then lead through the 
$\etat$ term above to small-scale helical fields
also with positive $\hf$. 
Moreover, because total helicity is conserved, the turbulent diffusion will
put a negative helicity $\hm$ on larger scales. Since this $\hm$ was
originally positive, this leads to a decrease of $\hm$. This
process may be behind the apparent forward cascade of helicity.

Now consider the large-scale field in the solar wind.
Realistically, this field will also have reversals
associated with the poloidal field from the Sun reversing every 11 years.
However, a unique sense (sign) of its helicity can be imprinted because the sense of rotation
of the Sun is always the same, and, in addition, the radial velocity is always outward.
We know from the work of \cite{Bie87,Bie87b}
that this sense is to give negative helicity
in the north, opposite to what the dynamo generated mean field had in the solar interior
and at the surface. Perhaps the differential rotation in the
solar wind pumps negative helicity to the north and positive to the south
so as to reverse the original sign of the mean field helicity.
One can then explain our current observations as follows:
First, the positive helicity seen at larger $k$ is that generated from the turbulent
diffusion spreading the helicity of the dynamo-generated mean field to larger $k$ and then advecting
this outwards.
Second, the Parker spiral eventually
leads to a negative helicity on the largest scales \citep{Bie87}.
In this picture, the helicity of the Parker field would correspond to scales
around $5\AU$ (corresponding to $k=1.2\AU^{-1}$
at $R<2.8\AU$; see \Fig{rmeanhel4b}).
This is also in agreement with early work of \cite{SB93} who found
negative magnetic helicity in the north below frequencies of about
$10\uHz$, corresponding to $k<20\AU^{-1}$ at a wind speed of about
$400\kms$.
However, this scenario needs to be much better explored through simulations
such as those of \cite{WBM11}.

Unfortunately, the relative magnetic helicity is rather weak.
Therefore, only through extensive averaging we are able to
extract any useful information.
This low level of relative magnetic helicity suggests that there is
efficient mixing taking place in the solar wind, but it could also mean
that the magnetic helicity is already rather low at the solar surface.
Another possible reason could be the proximity to the solar minimum
in 1996.

Although magnetic helicity fluxes are expected to keep their preferred
sign over the solar cycle, some modulation is definitely to be expected,
as was already found by \cite{SB93}
using measurements spanning the years 1965 to 1988.
The data span analyzed in the present work is too short to make any
meaningful statements, but one can see that at least at larger wavenumbers
there seems to be a decline in magnetic helicity as one progresses
further toward solar minimum in 1996.

The present data can be used to extract quantitative estimates for
magnetic helicity fluxes.
This quantity is normally quoted in Maxwell squared per solar cycle.
Assuming that the magnetic helicity during 1993--1996 is representative
of the rest of the solar cycle, our analysis suggests values around
$10^{45}\Mx^2/{\rm cycle}$, which is comparable to the earlier work
of \cite{Bie87b}.

Although the results presented in this paper are physically appealing,
there remains uncertainty
about the assumptions made in this work, most notably the Taylor hypothesis,
which may be problematic over  large length scales, but
it is expected to be reasonably well satisfied for fluctuations
in the inertial range of the turbulent spectrum \citep[see, e.g.,][]{MG82}.
The assumption of isotropy is only needed when computing three-dimensional
spectra.
Only the two components in the plane
perpendicular to the radial direction enter in our analysis
of magnetic helicity.
The finite pitch angle of the Parker spiral introduces anisotropy in that
plane \citep{MGOR096}.
This becomes important at progressively smaller length scales \citep{Wicks11},
even though the variance of the three components remains similar
(see \Fig{pdistance}).
Using spectra of the different components of the magnetic field,
\cite{Bie96} was able to determine that $\sim85\%$ of the magnetic
energy resides in the perpendicular field fluctuations, while for
{\it Ulysses}, \cite{Smi03} found that this value is closer to $50\%$.
In any case, it would be useful to estimate magnetic helicity
using synthetic data from a numerical simulation. Such an exercise
is likely to provide useful insight into the reliability of the
assumptions made and the accuracy of the method.

\acknowledgments

AB and KS acknowledge the hospitality of the International Space Science
Institute in Bern, where this work was started.
This work was supported in part by
the European Research Council under the AstroDyn Research Project No.\ 227952
and the Swedish Research Council Grant No.\ 621-2007-4064.

\appendix

\section{Derivation of Equations (10) and (11)}
\label{Deriv}

We begin with \Eq{Relation1D3D}, but instead of performing the
integration over $k_T$ and $k_N$ we use cylindrical coordinates
$(k_\perp,\phi_k)$ in Fourier space and assume axisymmetry, so
therefore $\dd k_T\dd k_N=2\pi k_\perp\dd k_\perp$.
Thus, we have
\EQ
M_{ij}^{\rm1D}(k_R)
=\int_0^{2\pi}\int_0^\infty
M_{ij}^{\rm3D}(k_R,k_\perp,\phi_k)\,k_\perp\dd k_\perp\dd\phi_k
=2\pi\int_0^\infty M_{ij}^{\rm3D}(k_R,k_\perp)\,k_\perp\dd k_\perp.
\label{Relation1D3Daxi}
\EN
Next, we insert \Eq{IsotropicTensor}, take the trace,
use $k^2=k_R^2+k_\perp^2$, 
and substitute $k_\perp\dd k_\perp=k\,\dd k$ valid for a fixed $k_R$, 
to carry out the
integration over $k$ in the allowed range from $k_R$ to $\infty$, and
obtain, using \Eq{EM1D},
\EQ
\delta _{ij}M_{ij}^{\rm1D}(k_R)
=2\pi\int_{k_R}^\infty 2\,{2\mu_0 E_{\rm M}^{\rm3D}(k)\over8\pi k^2} \,k\,\dd k
=\int_{k_R}^\infty {\mu_0 E_{\rm M}^{\rm3D}(k)\over k}\, \dd k
=\int_{k_R}^\infty \mu_0 E_{\rm M}^{\rm3D}(k) \, \dd\ln k,
\label{Relation1D3Daxi2}
\EN
which corresponds to \Eq{EMto1D}.
Likewise, multiplying instead with $-2\epsilon_{ijR}\ii k_R^{-1}$,
and using \Eq{HM1D}, we arrive at
\EQ
2\epsilon_{ijR}(\ii k_R)^{-1} M_{ij}^{\rm1D}(k_R)
=2\pi\int_{k_R}^\infty 4\,{H_{\rm M}^{\rm3D}(k)\over8\pi k^2} \,k\,\dd k
=\int_{k_R}^\infty {H_{\rm M}^{\rm3D}(k)\over k}\, \dd k
=\int_{k_R}^\infty H_{\rm M}^{\rm3D}(k) \, \dd\ln k,
\label{Relation1D3Daxi3}
\EN
which corresponds to \Eq{HMto1D}.
In particular, this yields $\mu_0 E_{\rm M}^{\rm1D}(k_R)=|\hat{\BB}|^2$
and $H_{\rm M}^{\rm1D}(k_R)=4\,{\rm Im}(\hat{B}_T\hat{B}_N^\star)/k_R$,
as stated just below \Eq{M1D}.


\begin{thebibliography}{}

\bibitem[Balogh et al.(1992)]{Bal92}
Balogh, A., Beek, T. J., Forsyth, R. J., Hedgecock, P. C., Marquedant, R. J., Smith, E. J., Southwood, D. J., \& Tsurutani, B. T.\yanas{1992}{92}{221}

\bibitem[Bao et al.(1999)]{Bao99}
Bao, S. D., Zhang, H. Q., Ai, G. X., \& Zhang, M.\yanas{1999}{139}{311}

\bibitem[Berger(1997)]{Ber97}
Berger, M. A.\yjgr{1997}{102}{2637}

\bibitem[Berger \& Ruzmaikin(2000)]{BR00}
Berger, M. A., \& Ruzmaikin, A.\yjgr{2000}{105}{10481}

\bibitem[Bieber et al.(1987a)]{Bie87}
Bieber, J. W., Evenson, P. A., \& Matthaeus, W. H.\yapj{1987a}{315}{700}

\bibitem[Bieber et al.(1987b)]{Bie87b}
Bieber, J. W., Evenson, P. A., Matthaeus, W. H.\yjgr{1987b}{14}{864}

\bibitem[Bieber \& Rust(1995)]{BR95}
Bieber, J. W. \& Rust, D. M.\yapj{1995}{453}{911}

\bibitem[Bieber et al.(1996)]{Bie96}
Bieber, J. W., Wanner, W., \& Matthaeus, W. H.\yjgr{1996}{101}{2511}

\bibitem[Blackman \& Brandenburg(2003)]{BB03}
Blackman, E. G., \& Brandenburg, A.\yapjl{2003}{584}{L99}

\bibitem[Blackman \& Field(2000)]{BF00}
Blackman, E. G., \& Field, G. B.\ymn{2000}{318}{724}

\bibitem[Brandenburg(2001)]{B01}
Brandenburg, A.\yapj{2001}{550}{824}

\bibitem[Brandenburg(2009)]{Bra09}
Brandenburg, A.\yjour{2009}{Plasma Phys. Control. Fusion}{51}{124043}

\bibitem[Brandenburg et al.(2009)]{BCC09}
Brandenburg, A., Candelaresi, S., \& Chatterjee, P.\ymn{2009}{398}{1414}

\bibitem[Brandenburg \& Matthaeus(2004)]{BM04}
Brandenburg, A., \& Matthaeus, W. H.\ypre{2004}{69}{056407}

\bibitem[Brandenburg \& Sandin(2004)]{BS04}
Brandenburg, A., \& Sandin, C.\yana{2004}{427}{13}

\bibitem[Brandenburg \& Subramanian(2005)]{BS05}
Brandenburg, A., \& Subramanian, K.\yjour{2005}{Phys.\ Rep.}{417}{1}

\bibitem[Brandenburg et al.(2003)]{BBS03}
Brandenburg, A., Blackman, E. G., \& Sarson, G. R.\yjour{2003}{Adv. Space Sci.}{32}{1835}

\bibitem[Chae(2000)]{Chae00}
Chae, J.\yapj{2000}{540}{L115}

\bibitem[Choudhuri et al.(2004)]{CCN04}
Choudhuri, A. R., Chatterjee, P., \& Nandy, D.\yapj{2004}{615}{L57}

\bibitem[D\'emoulin \& Berger(2003)]{DB03}
D\'emoulin, P., \& Berger, M. A.\ysph{2003}{215}{203}

\bibitem[D\'emoulin et al.(2002)]{Dem02}
D\'emoulin, P., Mandrini, C.~H., van Driel-Gesztelyi, L.,
Thompson, B.~J., Plunkett, S., Kov\'ari, Z., Aulanier, G.,
\& Young, A.\yapj{2002}{382}{650}

\bibitem[Dikpati \& Gilman(2001)]{DG01}
Dikpati, M., \& Gilman, P. A.\yapj{2001}{559}{428}

\bibitem[Dobler et al.(2003)]{Dobler_etal03}
Dobler, W., Haugen, N. E. L., Yousef, T. A., \& Brandenburg, A.\ypre{2003}{68}{026304}

\bibitem[Freeman(1998)]{Fre98}
Freeman, J. W.\ygrl{1998}{15}{88}

\bibitem[Goldstein et al.(1991)]{GRF91}
Goldstein, M. L., Roberts, D. A., \& Fitch, C. A.\ygrl{1991}{18}{1505}

\bibitem[Goldstein et al.(1995)]{GRM95}
Goldstein, M. L., Roberts, D. A., \& Matthaeus, W. H.\yaraa{1995}{33}{283}

\bibitem[Ji(1999)]{Ji99}
Ji, H.\yprl{1999}{83}{3198}

\bibitem[Keinigs(1983)]{Kei83}
Keinigs, R. K.\ypf{1983}{26}{2558}

\bibitem[Kleeorin et al.(2000)]{KMRS00}
Kleeorin, N., Moss, D., Rogachevskii, I., \& Sokoloff, D.\yana{2000}{361}{L5}

\bibitem[Kusano et al.(2002)]{Kus02}
Kusano, K., Maeshiro, T., Yokoyama, T., \& Sakurai, T.\yapj{2002}{577}{501}

\bibitem[Matthaeus et al.(1996)]{MGOR096}
Matthaeus, W. H., Ghosh, S., Oughton, S., Roberts, D. A.\yjgr{1996}{101}{7619}

\bibitem[Matthaeus \& Goldstein(1982)]{MG82}
Matthaeus, W. H., \& Goldstein\yjgr{1982}{87}{6011}

\bibitem[Matthaeus et al.(1986b)]{MGS86b}
Matthaeus, W. H., Goldstein, M. L. \& King, J. H.\yjgr{1986b}{91}{59}

\bibitem[Matthaeus et al.(1986a)]{MGS86}
Matthaeus, W. H., Goldstein, M. L. \& Lantz, S. R.\ypf{1986a}{29}{1504}

\bibitem[Matthaeus et al.(1982)]{MGS82}
Matthaeus, W. H., Goldstein, M. L., \& Smith, C.\yprl{1982}{48}{1256}

\bibitem[Moffatt(1969)]{Mof69}
Moffatt, H. K.\yjfm{1969}{35}{117}

\bibitem[Narita et al.(2010)]{NSGG10}
Narita, Y., Sahraoui, F., Goldstein, M. L., \& Glassmeier, K. H.\yjgr{2010}{115}{A04101}

\bibitem[Parker(1958)]{Par58}
Parker, E. N.\yapj{1958}{128}{664}

\bibitem[Pevtsov et al.(1995)]{Pev95}
Pevtsov, A. A., Canfield, R. C., \& Metcalf, T. R.\yapj{1995}{440}{L109}

\bibitem[Pevtsov \& Latushko(2000)]{Pev00}
Pevtsov, A. A. \& Latushko, S. M.\yapj{2000}{528}{999}

\bibitem[Rust \& Kumar(1994)]{RK94}
Rust, D. M. \& Kumar, A.\ysph{1994}{155}{69}

\bibitem[Rust \& Kumar(1996)]{RK96}
Rust, D. M. \& Kumar, A.\yapjl{1996}{464}{L199}

\bibitem[Sahraoui et al.(2010)]{Sah10}
Sahraoui, F., Goldstein, M. L., Belmont, G., Canu, P., \& Rezeau, L.\yprl{2010}{105}{131101}

\bibitem[Sahraoui et al.(2009)]{Sah09}
Sahraoui, F., Goldstein, M. L., Robert, P., \& Khotyaintsev, Y. V.\yprl{2009}{102}{231102}

\bibitem[Seehafer(1990)]{See90}
Seehafer, N.\ysph{1990}{125}{219}

\bibitem[Seehafer(1996)]{See96}
Seehafer, N.\ypre{1996}{53}{1283}

\bibitem[Smith(2003)]{Smi03}
Smith, C. W.\yproc{2003}{413}
{AIP Conf.\ Proc.\ 679, Proc.\ Tenth International Solar Wind Conference}
{M.\ Velli, R.\ Bruno, F.\ Malara, \& B.\ Bucci}
{Melville, NY: AIP}

\bibitem[Smith \& Bieber(1993)]{SB93}
Smith, C. W., \& Bieber, J. W.\yproc{1993}{493}
{23rd International Cosmic Ray Conference, Vol. 3}
{D. A. Leahy, R. B. Hicks, and D. Venkatesan}
{Singapore: World Scientific}

\bibitem[Stribling et al.(1995)]{SMO95}
Stribling T., Matthaeus W. H., \& Oughton S.\ypp{1995}{2}{1437}

\bibitem[Subramanian \& Brandenburg(2006)]{SB06}
Subramanian, K., \& Brandenburg, A.\yapj{2006}{648}{L71}

\bibitem[Tu \& Marsch(2003)]{TM95}
Tu, C.-Y., \& Marsch, E.\yssr{1995}{73}{1}

\bibitem[Smith et al.(2001)]{Smith01}
Smith, C. W., Matthaeus, W. H., Zank, G. P., Ness, N. F., Oughton, S., \& Richardson, J. D.\yjgr{2001}{106}{8253}

\bibitem[Tennekes \& Lumley(1972)]{TL72}
Tennekes, H., \& Lumley, J. L.\ybook{1972}{First Course in Turbulence}
{Cambridge: MIT Press}

\bibitem[Warnecke et al.(2011a)]{WBM10}
Warnecke, J., Brandenburg, A., \& Mitra, D.\yproc{2011a}{arXiv:1011.4299}
{Advances in Plasma Astrophysics}
{A. Bonanno et al.}{IAU Symp. 274}

\bibitem[Warnecke et al.(2011b)]{WBM11}
Warnecke, J., Brandenburg, A., \& Mitra, D.\sana{2011b},
arxiv:1104.1613

\bibitem[Webb et al.(2010)]{Webb10}
Webb, G. M., Hu, Q., Dasgupta, B., \& Zank, G. P.\yjgr{2010}{115}{A10112}

\bibitem[Wicks et al.(2011)]{Wicks11}
Wicks, R. T., Horbury, T. S., Chen, C. H. K., \& Schekochihin, A. A.\yprl{2011}{106}{045001}

\bibitem[Yousef \& Brandenburg(2003)]{YB03}
Yousef, T. A., \& Brandenburg, A.\yana{2003}{407}{7}

\bibitem[Zhang et al.(2006)]{Zhang06}
Zhang, H., Sokoloff, D., Rogachevskii, I., Moss, D., Lamburt, V.,
Kuzanyan, K., \& Kleeorin, N.\ymn{2006}{365}{276}

\end{thebibliography}
\end{document}